\documentclass{aa}
\usepackage{graphicx}
\usepackage{times}
\begin{document}

\def\kms{\mbox{km\,s$^{-1}$}}
\def\Hubble{\mbox{km\,s$^{-1}$\,Mpc$^{-1}$}}
\def\Doppler{\mathcal{D}}
\def\lsim{\raisebox{-.5ex}{$\;\stackrel{<}{\sim}\;$}}
\def\gsim{\raisebox{-.5ex}{$\;\stackrel{>}{\sim}\;$}}
\def\Snutspace{$(S,\nu,t)$-space}
\def\lgSnutspace{$(\lg S,\lg \nu,\lg t)$-space}
\newcommand{\mrm}[1]{\mathrm{#1}}
\newcommand{\dmrm}[1]{_{\mathrm{#1}}}
\newcommand{\umrm}[1]{^{\mathrm{#1}}}
\newcommand{\Frac}[2]{\left(\frac{#1}{#2}\right)}
\newcommand{\eqref}[1]{Eq.~(\ref{#1})}
\newcommand{\eqsref}[2]{Eqs~(\ref{#1}) and (\ref{#2})}
\newcommand{\eqssref}[2]{Eqs~(\ref{#1}) to (\ref{#2})}
\newcommand{\figref}[1]{Fig.~\ref{fig:#1}}
\newcommand{\tabref}[1]{Table~\ref{tab:#1}}
\newcommand{\secref}[1]{Sect.~\ref{sec:#1}}

%\thesaurus{00(0;0;0)}
\title{Cascades of shocks in Active Galactic Nuclei and their radiation}
\titlerunning{Cascades of shocks in AGN}
\author{
T.J.-L. Courvoisier \inst{1,2}  and M. T\"urler \inst{1,2}
}
\institute{
\textit{INTEGRAL} Science Data Centre, ch. d'\'Ecogia 16, CH-1290 Versoix, Switzerland \and
Geneva Observatory, ch. des Maillettes 51, CH-1290 Sauverny, Switzerland
}
\offprints{T. Courvoisier (ISDC)}
\mail{Thierry.Courvoisier@obs.unige.ch}
\date{Received /Accepted 11/08/05}

\abstract{
We discuss accretion flows on massive black holes in which different elements of the flow (clumps) have velocities that may differ substantially. We estimate the consequence of collisions between these clumps as they come close to the central object and calculate the resulting radiation. We show that this radiation is similar to that observed in the optical to X-ray spectral domain in Seyfert galaxies and quasars.  We also show that the large scale accretion is likely to be clumpy when arriving in the active region and that the clumps keep their identity between collisions.
}

\maketitle

\section{Introduction}
%\label{sec:introduction}

Although the basic elements of the physics of Active Galactic Nuclei (AGNs) were established soon after the discovery of quasars in the early 1960's and are now beyond dispute, many aspects of their physics still have to be discovered. We know that AGNs are powered by the gravitational energy of matter that falls into the deep gravitational well created by  massive black holes, and  can estimate the mass of the black holes and the accretion rates (or at least lower limits to these quantities) from the bolometric luminosity of the objects. The masses of the black holes may also be deduced from kinematic measurements (e.g. \cite{kaspietal00}, \cite{petersonetal04}, \cite{paltaniturler05}).  We have also identified many of the emission processes that create the continuum spectral energy distributions and the emission lines. We can  deduce the state of some of the emission gas from line ratios and the shape of the continuum emission. These facts are now well documented in a number of books and reviews (see e.g. \cite{SettiSwings2001}). 

While our understanding of the radiative processes that cool the electrons is very advanced, our understanding of the processes that transfer energy from the accreting mass to the radiating electrons is still quite unsatisfactory. In a stationary situation, the electron cooling rate  must be equal to the electron heating rate. There are several ways to transfer energy to electrons. Electric fields do so. It is difficult, however, to see how organized electric fields could arise across 100 $R_S$ ($R_S = \frac{2GM}{c^2}$, $M$ being the mass of the black hole) from the energy of the accreting ions.  Coulomb collisions are also a process in which energy from ions is transferred to electrons. We will see in Sect.~\ref{sec:OTS} one case where this process  plays an important role.

Optically thick and geometrically thin accretion disks (\cite{ShakuraSunyaev73}) have been used widely either to understand accretion and the corresponding radiation of the gravitational energy in the blue bump of AGNs or as support of more complex structures including Comptonising coronae (\cite{Haardtetal94}). This continues despite the  difficulties met by  accretion disk models  in  accounting for the  variability in the blue bump of AGN (see Sect.~\ref{sec:obsfacts}) and the lack of accepted sources of viscosity in the disk material large enough to remove angular momentum at the needed rate (\cite{koratkarblaes99}).

%Much of this understanding is based on the standard accretion disk model (\cite{ShakuraSunyaev73}) although this structure is known to be based on an unspecified viscosity and to have a number of difficulties when confronted with the observations (see Sect.~\ref{sec:obsfacts}.

In the following we propose  a framework in which the accreting  flow is much more structured and clumpier than a standard accretion disk, and we then make explicit use of the clumpiness of the flow to explain the observed properties of AGN.  We  consider a flow in which shocks are the primary means through which the accreting mass  transfers its kinetic energy to the radiating electrons.   We find that cascades of shocks are expected, first in an optically thick environment and then in  optically thin conditions. We  estimate the properties of the radiation emitted and show that they  match those observed in the optical to X-ray emission of AGNs. We can also understand the origin of relativistic electrons in this framework. However, the presence of very organized relativistic outflows (jets) does not follow from the shocks we discuss, but requires an additional phenomenology, probably based on the rotation of the central black hole.

We consider  accretion flows characterised by very inhomogeneous conditions and use the fact that  shocks in such a flow are an efficient way to transform bulk kinetic energy in thermal and non-thermal relativistic particles. We then  calculate the resulting radiation field.

In Sect.~\ref{sec:obsfacts} we summarize the most prominent observational evidence on AGN that are relevant for our discussion. We then present the main features of the model and the nature of the accretion flow in Sect.~\ref{sec:casshocks}.  In Sect.~ \ref{sec:OthickS} and Sect.~ \ref{sec:OTS} we describe quantitatively the shocks in the accretion flow that energetize the electrons and discuss the resulting electromagnetic radiation. 

\section{Some observational results}
\label{sec:obsfacts}

We summarize here some observational results, mainly based on 3C~273, emphasizing those aspects which are difficult to understand in the frame of accretion disk models with or without a corona. 

\subsection{Blue bump}

Seyfert galaxies and quasars have a broad optical-UV emission that is markedly in excess of the extrapolation of the infrared emission to high frequencies. This component, the blue bump, has been associated with the accreting matter since \cite{shields78}. This emission is broader than a single black body, and has been modeled as an optically thick and geometrically thin accretion disk by \cite{malkanseargent82} and many authors since then. This component has been associated with shocks in an outflowing wind by \cite{camenzindcourvoisier83}. Despite many years of efforts, the blue bump has still not been  described well by a standard optically thick and geometrically thin accretion disk, even when several effects that may distort the original disk emission are taken into account (\cite{koratkarblaes99}). 

The shape of the blue bump, and in particular the energy at which it is observed, is very similar in objects of very different luminosities (\cite{walteretal94}).  Accretion disk  models predict, however, that the temperature of the disc depends on the mass of the object, hence on its luminosity. The lack of correlation between the shape of the blue bump emission and the object luminosity  can thus  be compatible only with standard accretion models if the mass of the black hole and the rate of accretion are in a finely tuned relation (\cite{walterfink93}).

It has been known for a long time (\cite{CourvoisierClavel91} and references therein) that the blue bump varies so that there is at most a very short lag between the UV and the optical emission. This suggests that the information about the flux variations travels at speeds close to the speed of light between the hotter and the cooler parts of the blue bump emission region rather than at sound or viscous velocities. 

The optical and ultraviolet variability of 3C~273 has been studied by \cite{Paltanietal98}. It was found that the  lag between the 1250\,\AA\  and long wavelength lightcurves,  $\Delta_t(\lambda)$, depends on the wavelength in the following way:

\begin{equation}
\label{eq:lag}
\Delta_t(\lambda) \simeq 4\,\mbox{days} \cdot \log_2(\lambda/1275\mbox{\AA})  .
\end{equation}

%The cross correlations between the optical and UV lightcurves also show systematically a secondary maximum at 1.8\,year.  
\noindent
A study of NGC~7469 by \cite{wandersetal97} and \cite{collieretal98} also showed  a delay increasing with wavelength and suggested that the delay in this object is  about 0.6\,days rather than 4 days as in 3C~273.

It has been shown by \cite{Paltanietal98} that the UV light curve of 3C~273 can be described by the superposition of independent finite duration events. The available data constrain  the shape  of the individual events only  crudely, which can be characterized by a linear  increase up to a maximum luminosity followed by a linear  decrease. The  increase and decrease durations may be assumed to be the same. This parametrization is also used in the following. It will be shown that the details of the event shape do not influence the conclusions that can be reached. The rate of events was found to be about 3 per year, their duration $2\mu$ ($\mu$ being the time needed to reach peak intensity) to be about 0.4\,year, and the energy of each event to be on the order of $10^{52}$\,erg.

\subsection{X-ray - blue bump correlations}

The X-ray slope in the 2-10 keV domain of 3C~273 is about 0.5 and shows some variations. INTEGRAL observations in January 2003 showed that at this epoch the hard X-ray emission (2-200\,keV) was at one of the weakest historical levels and that the slope was considerably steeper than at earlier epochs (\cite{courvoisieretal03}).

The short lags between the light curves of the blue bump emission at different wavelengths has most often been explained by assuming that the gravitational energy released by the accreting matter is not dissipated in the optically thick accretion disk, but instead in a corona outside the disk that illuminates it and heats it. In this way, variations in the heating  (hard X-ray) flux indeed causes  quasi-simultaneous variations in the different parts of the blue bump emission material. This model predicts that the heating flux and the reprocessed blue bump emission are tightly correlated at zero lag. Measurements of the heating flux are difficult, as most of the heating energy is radiated at energies higher than 10 keV, whereas most of the X-ray photons are emitted at energies less than 10\,keV. Most of the observations were therefore performed at energies less than 10\,keV. Observations of the quasar 3C~273 in the hard X-rays were performed with the BATSE instrument on CGRO and with INTEGRAL, they  do not show the expected correlation  with the blue bump emission(Fig.~1).

\begin{figure*}[tb]
\includegraphics[width=12cm]{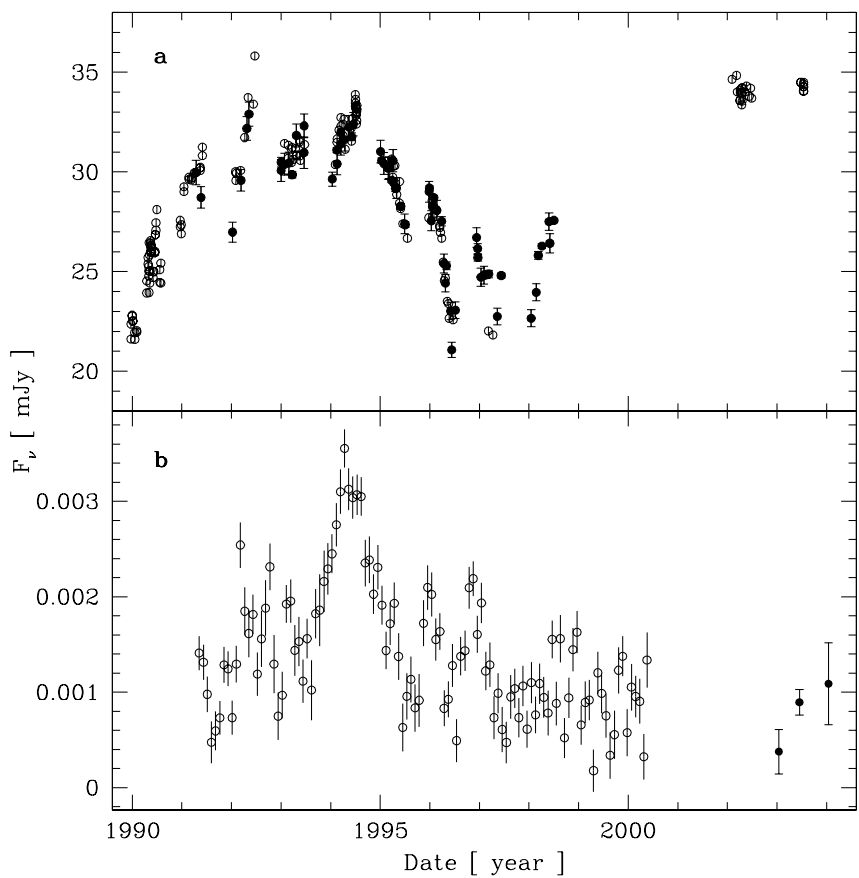}
\hfill
\parbox[b]{55mm}{
\caption{\label{fig:lc}
Comparison of \textbf{a)} the optical B band lightcurve and \textbf{b)} the 60
keV X-ray lightcurve of 3C 273. Optical data are from the Geneva photometric
system (open points) and scaled observations from Kaspi et al. (2000) (filled
points). X-ray observations are from CGRO/BATSE (averaged into 30-days bins)
(open points) and from INTEGRAL/ISGRI (filled points).}}
\end{figure*}

Studying the cross correlation between the X-ray and the UV lightcurves of 3C~273, \cite{Paltanietal98b} observed that the cross correlation has two relative maxima, one at zero lag and the other suggesting that the UV emission leads the X-ray emission by approximately 1.8\,years. This result is based on  observation of one major event that was well covered by both light curves, it will therefore need confirmation when many such events can be studied simultaneously.

\subsection{Line emission}

Quasars and other AGNs (with the notable exception of BL Lac objects) show prominent line emission. The line emitting material  has a density of about $10^{10}$\,cm$^{-3}$, and is photo-ionized (see e.g. \cite{netzer90}).  This line emission material fills only a small fraction of the volume, it has been discussed mainly in terms of isolated clouds. The clouds have not yet been observed, nor has the discrete nature of the line emission been established. \cite{dietrichetal99}  attempted to measure the signature that the presence of discrete clouds should leave on the line profile in 3C~273. Even at very high resolution and high signal-to-noise, the line profiles are smooth. Only lower limits to the number of clouds contributing to the lines  could be found. The number of individual clouds in the broad line region of 3C~273 is thus larger than $\sim 10^8$. This very high number may imply that a description in terms of individual clouds  may not be appropriate, but that a better model may be found in a highly disorganized  matter flow.

\section{Cascades of shocks}
\label{sec:casshocks}

 In order to explain the complex properties of the emission of AGN outlined above, we discuss the emission expected from an accretion flow that is characterized by individual clumps rather than a smooth flow. The flow generates  first optically thick shocks, then optically thin ones. These shocks  are are the origin of the radiation.
We consider a clumpy flow in which the clumps have velocities  given by the gravitational  field of the central black hole and a wide distribution of velocity directions. Clumps following orbits in the  gravitational field of the central black hole will lose energy either through tidal processes in the vicinity of the black hole or when they interact with one another. Only the latter process can dissipate the bulk kinetic energy of the clumps and thus be at the origin of the large luminosities of AGNs. We therefore consider in the following  the shocks that form at the boundary of two colliding clumps. We then follow the shocked material as it interacts with surrounding material having also been shocked in similar conditions. We show that the radiation emitted as a consequence of these processes closely matches the observed optical to X-ray properties of AGNs.

The phenomenology we describe here relies on the existence of a very clumpy accretion flow.  This is expected on at least two grounds (i)  the interstellar matter from which accretion in AGNs is nurtured is very structured and certainly not smooth; (ii)  self gravity in the accretion flow is likely to play an important role at distances larger than 1\,pc from the black hole (see \cite{collinzahn99} and references therein).  These authors discuss how stars of about 10 solar masses form in the 1-100\,pc region of an accretion disk and how they migrate to the central regions. The mass range of stars that are being produced does correspond to the clumps  at the origin of the phenomenology described here.

In past years \cite{Courvoisieretal96} and \cite{Torricellietal00} have proposed and discussed a model in which the accreted gas clumps were assumed to be stars. Although it may well be that a fraction of the accreted matter is indeed in the form of stars, the approach taken here does not rely on this assumption. It is sufficient to request that the primary shocks be optically thick.  The nature of the clumps can thus be very diverse, from stars (be they from the host galaxy as in our previous work or generated in the accretion flow), to dense  elements of an accretion disk. The clumps may also form  in the accreted interstellar medium as discussed in Sect. \ref{sec:clouds} or even  in outflowing winds.  Some properties of the clouds will be discussed  in Sect. \ref{sec:clouds}. For shocks to form, it is mandatory that clumps have velocities that differ either in direction or magnitude so that the relative velocities are  the same order as the clump velocities. This  suggests that clumps have  a broad distribution of angular momentum, as  expected if the accreted matter comes from regions that have been disturbed by either non-spherical potentials or from merging and/or interacting galaxies. The colliding clumps may then be thought of as belonging to two (or possibly a distribution of) flows, each characterized by its specific angular momentum and origin.  We consider here shocks between "generic clumps". The results obtained will then be used to develop more realistic approaches to AGN including a discussion of distributions of clump properties like mass, size  and angular momentum, and a discussion of the properties of the region surrounding the black hole like density  and velocity distributions.

\section{Optically thick shocks}
\label{sec:OthickS}

The energy content in the outbursts of the blue bump of the bright quasar 3C~273 has been calculated by \cite{Paltanietal98} to be $\simeq 10^{52}$\, ergs . This is the kinetic energy of some $10^{33}$\,g at $0.1\, c$, the Kepler velocity at 100 $R_S$. Since not all the kinetic energy is radiated, we  propose that typical shocks involve clumps on the order of several $10^{33}$\,g. Such masses  agree with the discussion in  Sect.~\ref{sec:casshocks}.

We therefore consider shocks between clumps of mass $M=M_{33}10^{33}$\,g  and size $R_\mathrm{c}$ moving  at $v_\mathrm{Kepler}$ in the gravitational field of the central black hole. $M_{33}$ will typically range from a few to a few 10s. Around 100\,$R_S$, $v_\mathrm{Kepler} = \beta_{0.1}\,0.1 c$ with $\beta_{0.1}$  of order 1.  We assume the clumps to be sufficiently dense  for the photon diffusion time $t_\mathrm{diff}$ within the matter to be very long compared to the crossing time $t_\mathrm{cross} = \frac{R_\mathrm{c}}{v_\mathrm{Kepler}}$. We will consider optically thin shocks in  sect. \ref{sec:OTS}.

When two such clumps  collide, their kinetic energy in the center of mass frame is thermalized. The resulting configuration is a very hot dense gas cloud in adiabatic expansion.  This is similar to a supernova in many ways. The properties of the expansion are therefore expected to resemble those of expanding supernovae in their early stages. The total energy of the gas cloud in its rest frame, that of the center of mass of the two colliding clumps, is $E\,=\,M\,v_\mathrm{Kepler}^2$\,$\eta_\mathrm{geom}$, where  $\eta_\mathrm{geom}$ is a factor taking the geometry of the collision into account and is assumed to be 1 in the following.  Coulomb collisions at the origin of the thermalization being elastic, we expect that the expansion velocity of the gas cloud is $v_\mathrm{exp} \simeq$ $v_\mathrm{Kepler}$. The geometry of the expanding gas  reflects the symmetry of the collision. We assume in the following spherical expansion for the sake of simplicity. Detailed numerical simulations of stellar collisions (\cite{Freitag00}),  using smooth particle hydrodynamics methods and including the internal structure of the stars, confirm that  high velocity collisions between stars lead to the disruption of the stars and to a high velocity expanding gas cloud.

The expanding gas will radiate a fraction $\eta$ of its energy $E$ at its photosphere. The time it takes for this energy to be radiated can be estimated by comparing the photon diffusion time $t_\mathrm{diff}$ with the expansion time. The photon diffusion time is

\begin{equation}
\label{eq:tdiff}
t_\mathrm{diff} \simeq \frac{l}{c}\,(\frac{R}{l})^2 = \frac{R^2}{c}\,\sigma_\mathrm{T}\,n_\mathrm{e} ,
\end{equation}

\noindent
where $l = (\sigma_\mathrm{T}\cdot n)^{-1}$ is the photon mean free path, $n_\mathrm{e}$ the electron density, $R$ the size of the expanding cloud, and $\sigma_\mathrm{T}$ the Thomson cross section. Use of the Thomson cross section is sufficient as it is expected that the shocks will completely ionize the material. The expansion time is $t_\mathrm{exp} = \frac{R}{v_\mathrm{exp}}$.

The expansion time and diffusion times are equal; i.e. $t_\mathrm{exp} = t_\mathrm{diff}$,  at $R = R_\mathrm{max}$, which may be expressed as
 
\begin{equation}
\label{eq:Rmax}
 R_\mathrm{max} = (\frac{v_\mathrm{exp}}{c}\,\sigma_\mathrm{T}\,\frac{M}{m_\mathrm{p}}\,\frac{3}{4\pi})^{1/2} = 3.1 M_{33}^{1/2}\beta_{0.1}^{1/2}\,10^{15}\,\mbox{cm} ,
\end{equation}

\noindent
where now $\beta_{0.1} = v_\mathrm{exp}/(0.1\, c)$. $R_\mathrm{max}$ depends on the expansion velocity and the mass of the clumps. This size is reached after a time $t_\mathrm{exp}(R_\mathrm{max})\,=\,\frac{R_\mathrm{max}}{v_\mathrm{exp}}$, which for two mass clumps at 100 $R_S$, i.e. with $v_\mathrm{Kepler} \simeq 0.1\,c$ is on the order of

\begin{equation}
\label{eq:texp}
t_\mathrm{exp} = 1.0\,M_{33}^{1/2}\beta_{0.1}^{-1/2}\,10^6\,\mbox{s}.
\end{equation} 

\noindent
This time is comparable to the typical rise time of an event $\mu \simeq 0.2$ years, as measured in the UV light curves of 3C~273 (see Sect.~\ref{sec:obsfacts}, \cite{Paltanietal98}) for masses of several $10^{33}$\,g.

The expanding matter is very optically thick. The luminosity of the expanding mass is therefore  given by the size of the photosphere and its temperature, which  depends on the rate of the heat transfer to the exterior. The calculation of this rate is beyond the scope of the present paper. We can, however, estimate the temperature of the photosphere by assuming that it expands with  $v_\mathrm{exp}$ and that its luminosity is given by the simple model of \cite{Paltanietal98}, in which the luminosity increases linearly until it  reaches a maximum at a time $t = \mu$ after the beginning of the outburst.   We identify $\mu$ with $t_\mathrm{exp}(R_\mathrm{max})$.
With $L_{\mu}$ giving the luminosity at $t = \mu$, one obtains for the time dependence of the temperature $T(t)$, where $t$ is the time elapsed since the collision between two clumps:

\begin{equation}
\label{eq:Tt}
T(t) = [\frac{L_{\mu}}{4\pi \mu v_\mathrm{exp}^2 t \sigma}]^{1/4} ,
\end{equation}

\noindent
where $\sigma$ is the Stefan-Boltzmann constant.

The time dependence of the temperature (Eq.~\ref{eq:Tt}) may be written as

\begin{equation}
\label{eq:Tt2}
T(t) =  T_{\mu}\cdot (\frac{\mu}{t})^{1/4},
\end{equation}

\noindent
with

\begin{equation}
\label{eq:T0}
T_{\mu} = [\frac{L_{\mu}}{4\pi \sigma \mu^2 v_\mathrm{exp}^2 }]^{1/4} = 1.1\times 10^4 (\frac{L_{44}}{M_{33}\beta_{0.1}})^{1/4} \mbox{K}
\end{equation}

\noindent
is the temperature reached at the photosphere at $t = \mu$ and $L_{44} = \frac{L}{10^{44}}$\,ergs\,s$^{-1}$. The event properties derived from the observations of 3C~273 by \cite{Paltanietal98} suggest that the luminosity $L_{\mu}$ is a few $10^{44}$\,erg\,s$^{-1}$.

This estimate is valid only provided that the temperature of the material in expansion is sufficient for the surface to emit at the blackbody limit. We can check that this is the case by estimating the temperature at the begining of the expansion and the adiabatic rate of decrease in the temperature. The initial temperature is given by the shock velocity ($\simeq 0.1 c$) and is of the order of $10^{10}$\,K. The temperature decrease is given by the adiabatic expansion of a radiation pressure dominated gas $T \sim \rho^{1/3}$. Thus even assuming an initial clump or shock size as small as that of a star ($10^{11}$\,cm), the gas expanding to $R_\mathrm{max}$ does not adiabatically cool below  the photosphere temperature estimated above. 

 The model proposed here is able to account  for the blue bump emission of AGNs very well,  the typical temperatures of which  are indeed  on the order of $10^4$\,K and dominates the optical UV emission.  The model also predicts a broader component   than a single black body, because several events will be co-existing in the nucleus at different stages of their evolution and therefore at different individual temperatures.  This matches the observations. Different objects are expected to have blue bumps at roughly the same wavelengths, as this is given by collision parameters and not by the properties of an accretion disk that depends on the mass of the central object and the accretion rate. Differences between objects will furthermore be small, because of the low power in which the event parameters enter Eq.~\ref{eq:Tt2}.
 These parameters are such that the energy radiated during the event is on the order of 10$^{52}$\,ergs.

%Observations of SN1987A  have shown that the expansion of the envelope is very inhomogeneous. It is therefore to be expected that the same may be true of the expanding clouds in AGNs. 

\subsection{Lags in the light curves}

Equation \ref{eq:Tt2}  can be inverted to estimate the time it takes for the temperature of an event to cool from $T_1 = T(t_1)$ to $T_2 = T(t_2)$. With $\Delta T$ =  $T_1 - T_2$ one obtains the following for  $\Delta T$ small compared to $T_1$

\begin{equation}
\label{eq:t2t1}
t_2-t_1 = \mu\,(\frac{T_{\mu}}{T_1})^4\,\frac{4\Delta T}{T_1}   ,
\end{equation}
 to the first order in $\Delta T/T$.

It therefore takes about $10^6$\,s to cool by 25\% at $T_{\mu}$. This time is very similar to the lag $\Delta_t(\lambda)$ measured by \cite{Paltanietal98} in 3C~273 between the visible and the UV light curves (see Eq.~\ref{eq:lag}). Furthemore, the time delay increases as the temperature difference increases, corresponding to the observation that the lag of lightcurves increases as the wavelength difference between the light curves  increases (Eq.~\ref{eq:lag}).

\section{Optically thin shocks}
\label{sec:OTS}

The gas emerging from the expanding regions we have just described moves ballistically  with bulk velocities $v_\mathrm{exp}$ close to the Kepler velocity at the corresponding radius. Since the gas expands  beyond $R_\mathrm{max}$,  it becomes  optically thin at times larger than $t_\mathrm{exp}(R_\mathrm{max})$ and fills a  volume  of $\simeq R_\mathrm{max}^3$  that is comparable to the region within 100\,R$_S$, even for massive central black holes.  With an event rate  of several per year, the  gas envelopes  of the clouds therefore overlap in the central regions. A large fraction of the expanding gas of any cloud will  be shocked again, this time, however,  in an optically thin environment. The relative velocities can be expected to be somewhat larger than the sum of the Kepler velocities at the originating optically thick shocks due to the fact that gas moving inward and gravitationally accelerated has a higher probability of interacting with other expanding clouds than the gas moving outward.

The temperatures reached at the optically thin shocks are

\begin{equation}
\label{eq:Ts}
kT = \frac{3}{16} \mbox{\={m}} v_\mathrm{rel}^2   .
\end{equation}

\noindent
Here,  $v_\mathrm{rel}$ is the relative speed of the expanding  clouds where they shock  and \={m} is the average mass of a particle (=1\,$10^{-24}$\,g for completely ionized matter). With  $v_\mathrm{rel}$ of about 0.1\,c, the temperatures obtained are on the order of 1\,MeV.

The temperature of the electrons is given by the equilibrium between their  cooling  and their heating rates  in the shocked material. The electrons cool predominately  through Compton cooling. They gain energy through Coulomb interactions with the protons and nuclei, which do not cool through Compton radiation (or at least do so much less than electrons). The Coulomb interactions of the electrons with the protons and nuclei  is a relatively slow process. One therefore expects that the electron temperature is less than that of the protons.  

The average electron energy can be estimated by considering the equilibrium between energy gain through Coulomb collisions and Compton losses. Compton losses are given by (e.g. \cite{RybickiLightman79}): 

\begin{equation}
\label{eq:LC}
L_\mathrm{Compton} = \frac{4}{3}\sigma_\mathrm{T}c\gamma^2\beta^2\cdot u_\mathrm{ph} ,
\end{equation}

\noindent
 $u_\mathrm{ph}$ being the photon energy density and the other symbols having their usual meaning. The photon energy density is the one generated by the optically thick shocks of luminosity $L_\mathrm{thick}$:

\begin{equation}
\label{eq:uph}
u_\mathrm{ph} \simeq  \frac{L_\mathrm{thick}}{4\pi R_\mathrm{max}^2\,c}  \simeq 29 \frac{L_{44}}{M_{33}\, \beta_{0.1}} \mbox{erg\,cm}^{-3}.
\end{equation}

The electrons are heated through Coulomb collisions with the hot protons and ions at a rate  $L_{Coulomb} = E_\mathrm{p}/t_\mathrm{C}$ where $E_\mathrm{p}$ is the proton kinetic energy ($\simeq \frac{1}{2} m_\mathrm{p} v_\mathrm{rel}^2 \simeq 1$\,MeV) and $t_\mathrm{C}$ is given by the standard Coulomb energy transfer time (see e.g. \cite{Benz93}):

\begin{equation}
\label{eq:tC}
t_\mathrm{C} = \frac{v_\mathrm{e}^5}{4A_\mathrm{d}\cdot v_\mathrm{p}^2}   
\end {equation}

\noindent
with

\begin{equation}
\label{eq:Ad}
A_\mathrm{d} = \frac{8\pi n e^4 \ln(\Lambda)}{m_\mathrm{e}^2} = 3.2\, 10^{27}\cdot n_{8}\cdot \frac{\ln(\Lambda)}{20}
\end{equation}

\noindent
in c.g.s. units. The density may be estimated by 

\begin{equation}
\label{eq:density}
n = \frac{M/m_\mathrm{p}}{4\pi/3 R_\mathrm{max}^3} \simeq 5.3\times 10^9M_{33}^{-1/2}\,\beta_\mathrm{exp,0.1}^{-3/2}\,\mbox{cm}^{-3}.
\end{equation}

Using the photon energy density from Eq.~(\ref{eq:uph}) and the  electron density from Eq.~(\ref{eq:density}),  the equilibrium $L_\mathrm{Compton} = L_{Coulomb}$  can be solved for the electron energy:

\begin{equation}
\label{eq:Te}
\frac{E_\mathrm{e}}{m_\mathrm{e}c^2} \simeq 0.5 M_{33}^{1/7}\beta_{0.1}^{-1/7}L_{44}^{-2/7}E_\mathrm{p,MeV}, 
\end{equation}
where $E_\mathrm{p,MeV}  = E_\mathrm{p}$/1MeV. This estimate gives an average electron energy of some 250\,keV, which is   less than the proton temperature, as expected. It depends with a low power on the parameters of the model. 

Shocks are well known for producing relativistic particles that contribute  several percent of the energy balance (see e.g. \cite{Kirk94} for a discussion of particle acceleration in shocks). This will also take place in the environment we have just described. In the presence of a magnetic field, possibly also accreted with the in-falling matter, these relativistic particles will produce synchrotron emission. 

\subsection{X-ray emission}

The Compton process leading to the cooling of the $\simeq 250$\,keV electrons will radiate X-rays. For a non relativistic electron population the spectrum is characterized by a power law of index (\cite{RybickiLightman79}) $\alpha = -\frac{\ln(\tau)}{\ln(A)}$, where $\tau$ is the optical depth of the region. Using the density of Eq.~(\ref{eq:density}) and $R_\mathrm{max}$ as the size, one can estimate that $\tau \simeq 3.6\beta_{0.1}^{-1}$. $A$ is the relative energy gain per Compton diffusion, which is given by  $A = 4\,kT_\mathrm{e}/m_\mathrm{e}c^2$ for non relativistic electrons. $kT_\mathrm{e}$ is the average energy of the electrons as estimated in Eq.~( \ref{eq:Te}). For the parameters of our model, $\alpha \simeq  -0.3$. This slope   agrees with the observed hard X-ray slope of 3C~273 of about -0.5, considering that the spherical geometry used to deduce the optical depth, and hence the spectral slope, is certainly a very rough estimate at this stage.

It can be seen from Eq.~(\ref{eq:Te}) that an increase in the blue bump photon energy density leads to a decrease in the electron temperature. This in term leads to a decrease in $A$ and to a steepening of the X-ray emission. This behavior corresponds to the observations of 3C~273 made in 2003, when the blue bump was found to be at a very high level and the INTEGRAL slope was one of the steepest ever observed (\cite{courvoisieretal03}).

%A cut-off is expected at $\simeq 3\,kT_\mathrm{e}$ at energies of some 750\,keV, where indeed observations indicate that this feature is observed (see e.g. \cite{DeluitCourvoisier02}). 

We can estimate the X-ray luminosity of the optically thin shocks  starting from the Compton luminosity $L_\mathrm{Compton}$ of a single non relativistic electron  (Eq.~\ref{eq:LC} ) with the photon energy density from Eq.~(\ref{eq:uph}) and the electron velocity deduced  from Eq.~(\ref{eq:Te}):

\begin{equation}
\label{eq:Le}
L_\mathrm{Compton} \simeq 7.7\,10^{-13} M_{33}^{-6/7}\beta_{0.1}^{-8/7}L_{44}^{5/7} \mbox{ ergs\,s}^{-1}. 
\end{equation}

\noindent
For a shock in which there are  $  M_{33}/m_\mathrm{p}$\,electrons the X-ray luminosity is 

\begin{equation}
\label{eq:LX}
L_X \simeq 4.6\,10^{44} M_{33}^{1/7}\beta_{0.1}^{-8/7}L_{44}^{5/7}  \mbox{ergs\,s}^{-1}.
\end{equation}

Time variations of the resulting X-ray emission are expected on many time-scales that reflect the inhomogeneities of the colliding gas. These inhomogeneities result from those of the original gas as well as from those, on larger scales, due to the brightening and fading of the corresponding optically thick shocks. 

  Correlations of the X-ray emission with blue bump light curves are expected at short lags,  as the soft photons  are  being Comptonised, and also  at longer lags, because the optically thin shocks at the origin of the hot electrons have evolved from the optically thick shocks that created UV emission. The lag in this second case is on the order of  several  months ($\geq t_\mathrm{exp}(R_\mathrm{max})$), the UV light curve leading the X-ray emission. This behavior corresponds well to the observations described in \cite{Paltanietal98b}, where it is shown that the cross correlation of the X-ray and UV light curves of 3C~273  has two maxima, one indicating that the UV light curve leads the X-ray emission by $\simeq 1.75$\,year and the other indicating a good correlation with small lags. 

In the reference frame linked with the central black hole, the shocked material will have an angular momentum that is the vector sum of that of the material of the two original gas clumps. The shocks thus provide a means of reducing the angular momentum per particle and thus will allow the shocked material to fall deeper in the potential well, where further shocks may be expected. These cascades serve as an effective viscosity in the accretion process, while they contribute to transforming the gravitational energy of the incoming nucleons to the radiating electrons.

\section{Line emission}
 
 The gas moving outward beyond R$_\mathrm{max}$,  rather than inward following the expansion of the optically thick shocked material described in Sect.~\ref{sec:OTS}, will be slowed down to velocities less than $0.1\,c$ by its ballistic motion. This material is expected to be very inhomogeneous and have a wide range of velocities. Its density is given by Eq.~(\ref{eq:density}), which corresponds to the densities expected in the broad line emission region. We therefore suggest that the line emitting gas is the result of the shocked material once it has expanded in the outer regions of the nucleus.

 \section{Clouds}
 \label{sec:clouds}
 
 The structure of the clouds and the rate at which they initially fall into the central region of the AGN
 are given by the conditions in the central regions of the host galaxy and the geometry of the gravitational field  which drives the accretion from the large galactic scales to the immediate vicinity of the black hole where the high energy phenomenology is generated.  
 
In order to investigate how clumpy the accreted matter is likely to be, we consider the large scale accretion flow around an object  accreting  matter with a radial velocity $v_{acc}(r) = \chi v_{ff}(r)$ at the Eddington rate, where $v_{ff}(r)$ is the free-fall velocity onto the black hole. Those clouds that form while falling into the central regions of the AGN are those  with a mass larger than the Jeans mass in the conditions of the accreted matter and that have had sufficient  time to collapse during the process. In order to roughly estimate the collapse time of the clouds, we calculate the free fall of the cloud on itself

\begin{equation}
\label{eq:cff}
t_{cc} = \sqrt{\frac{s^3_{cloud}}{2GM_{J}}},
\end{equation}   
 
\noindent 
where $s_{cloud}$ is the initial size of the cloud while at large   distances to the active region and 
 
 \begin{equation}
 \label{eq:mjeans}
  M_J = (\frac{\pi}{3})^{\frac{5}{2}} (\frac{5kT}{G})^{\frac{3}{2}}n ^{-\frac{1}{2}}m_H^{-2},
 \end{equation} 
 
 \noindent
 is the Jeans mass for a density of gas $n$ at temperature $T$ (see e.g. \cite{lang}). The temperature of the matter far from the active region has been measured in the far infrared and found to cover a range with lowest measured temperatures around 50\,K (\cite{pollettaetal00}), which we use here as representative of the  accreted material. The density of the accreted matter is given by the mass accretion rate $\dot{M}$, itself given by the Eddington luminosity and the accretion velocity $v_{acc}$. In a spherical geometry one finds:
 
 \begin{equation}
 \label{eq:clouddens}
 n = \frac{\frac{\dot{M}}{m_H}}{4\pi r^2v_{acc}} .
 \end{equation}
 
 \noindent
 This is likely to be a lower limit, as the large scale accretion flow is unlikely to be spherically symmetrical. The Jeans mass is thus expected to be smaller than estimated from Eq. \ref{eq:mjeans}.
 
  Using  Eddington luminosity, an  accretion efficiency of 10\%,  and a black hole mass of 10$^9$ solar masses, we show  the Jeans mass at various initial distances to the active region as a function of the parameter $\chi$ in Fig. \ref{fig:cc}. In the same figure  we show in grey the region of the parameter space in which the cloud collapse time  $t_{cc}$ is shorter than the accretion time.  Figure \ref{fig:cc} thus shows, that for likely initial distances of 10\,pc and more, clumps of 100 solar masses and more can form and collapse, provided that the ratio $\chi$ of the accretion velocity to free fall velocity is  of several percent or less.  
  
 Departure from the spherical symmetry, assumed here for simplicity, increases the density of the accreted matter and therefore decreases the Jeans mass at a given distance. Any initial clumpiness of the accreted flow also favors the formation of clouds. It is therefore very likely that matter reaching the inner regions of a quasar will indeed be very clumpy rather than smooth.

\begin{figure*}[tb]
\label{fig:cc}
\includegraphics[width=12cm]{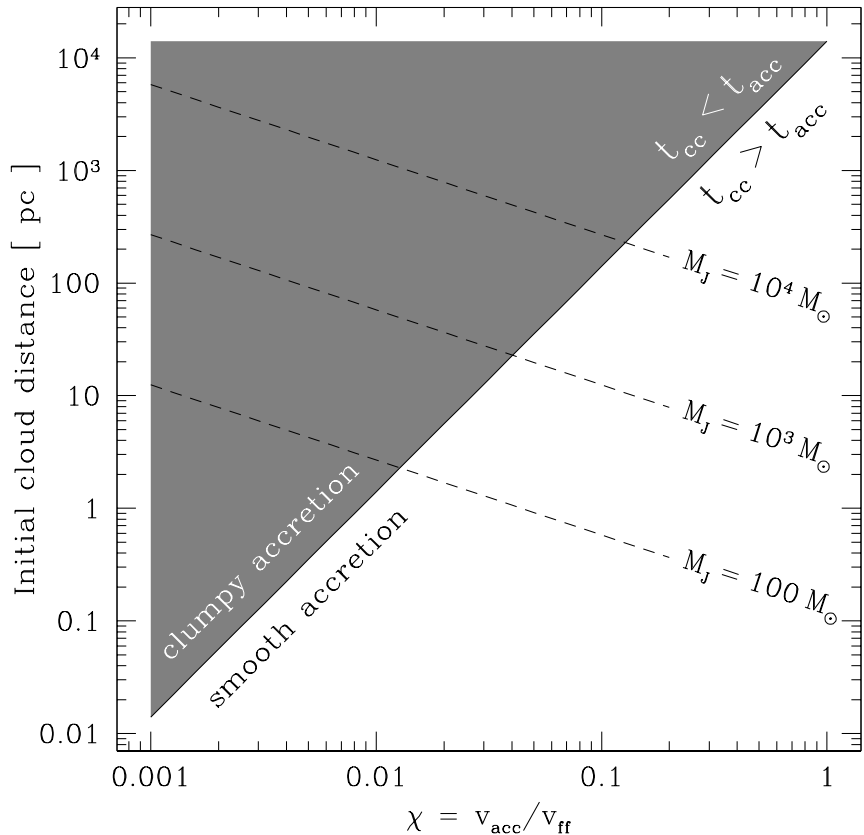}
\hfill
\parbox[b]{55mm}{
\caption{\label{fig:cc}
The Jeans mass as a function of the distance to the black hole and the ratio $\chi$ of the accretion velocity to free fall. Also shown in grey is the region in which cloud collapse is faster than accretion. Matter accreted from within the grey region of parameter space will be clumpy when arriving in the vicinity of the black hole, matter accreted from the white section of parameter space may be smooth, provided that it is homogeneous to start with.}}
\end{figure*}

% Self gravity is, however, likely to be an important property of the accreted clouds and  suggests that the mass of the clouds may be of the order of several solar masses (\cite{collinzahn99}), as required by the luminosity of individual events.
 
 Neglecting self gravity and the cloud structure, we may estimate the deformation of the clouds while in the active region in the typical time between collisions. Following \cite{Courvoisieretal96}, the number of clouds N in a region of size $R$ is given by the rate of collisions $\tau$:
 
 \begin{equation}
 \label{eq:Ncloud}
  N = \sqrt{\frac{4}{3}\frac{R^3}{R_c^2}\frac{1}{v_{Kepler}(R)}\tau} \simeq 20\frac{1}{R_{c,15}} \sqrt{\tau_{py}},
  \end{equation}
  
 \noindent 
  for a region of size $ R \simeq 100\times R_S \simeq 3\times 10^{16}$\,cm (corresponding to $M_{BH} = 10^9M_\odot$),  a cloud size $R_c = 10^{15}$\,cm, and $\tau_{py}$ collisions per year.
  
  With the time between two collisions given by $\frac{N}{\tau}$ and the shear generated by the Kepler velocity difference across the cloud one obtains that the cloud elongation during the time elapsed between 2 collisions along its orbit is about
  
  \begin{equation}
  \label{eq:deformation}
  \frac{\Delta R_c}{R_c} \simeq 20\frac{1}{R_{c15}}\frac{1}{\sqrt{\tau_{py}}}.
  \end{equation}
  
 \noindent 
 Since we neglected the self gravity of the clouds, this is an upper limit to the distortion of the clouds. Equation~(\ref{eq:deformation}) shows that, while the clouds may be significantly distorted, they are likely to keep their identity and remain isolated on their orbits between major collisions.

\section{Summary and conclusion}

\label{sec:discussion}

The picture that has been drawn is one of matter being accreted in the form of clumps of matter falling in the deep potential well of a massive black hole. These clumps interact with one another at first in the form of optically thick shocks and then, following an adiabatic expansion, through several optically thin shocks. The optically thick shocks give rise to the optical-UV emission, while the optically thin shocks give rise to the X-ray emission.

The approach taken here has several elements that are similar to what would be expected from AGN models based on supernovae (\cite{Aretxagaetal97}), the main difference being that the source of energy  is gravitational energy in the potential well of the central black hole rather than nuclear energy or  gravitational binding energy resulting from the collapse of single evolved stars. This then allows  the individual events to have properties like  total energy,  event  rate, or  event duration that all depend on their environment, typically the density of clumps in the surroundings of the black hole or their angular momentum distribution. This dependence  is important to understand the  time variability of AGN as a function of their average luminosity,  as discussed in \cite{Torricellietal00}, or the wide variety of observed properties seen in the class of Seyfert galaxies and quasars.

The model we have developed here allows us to understand many of the properties of AGN, namely: 
\begin{itemize}
\item
the temperature of the blue bump,
\item
the small lag observed between the UV and optical AGN light curves,
\item
the luminosity and shape of the hard X-ray emission,
\item
the correlations between the blue bump and X-ray light curves,
\item
the origin,  velocity, and  density of the broad line emission clouds.
\end{itemize}  
Moreover, the model does not make reference to any unspecified source of viscosity.

 AGNs have blue bumps that differ considerably in importance relative to other components of their continuum energy distribution. In the picture proposed here, this is related to the  luminosity of the optically thick shocks compared to that of the optically thin shocks. It is indeed expected  that a fraction of the matter (e.g. that which is only weakly bound in the accreted clouds) is  not shocked first in optically thick conditions. Objects for which this fraction is large have weak blue bumps. The temperature of the blue bump remains, however, similar, as it is given by the Kepler velocity.

There is one element of AGN physics that does not follow from the phenomenology we have considered, the existence of structured relativistic jets. Their generation does rely on further elements such as the rotation of the central black hole and magnetic fields.

%\begin{figure*}[tb]
%\includegraphics[width=12cm]{/home/astrox/turler/3Cfit/graph/geometry.eps}
%\hfill
%\parbox[b]{55mm}{
%\caption{\label{fig:geometry}%
%Sketch of the jet geometry discussed in \secref{model}, as observed in the rest frame of the quasar.
%We assume the contribution of the inner jet to be negligible with respect to }}
%\end{figure*}

\begin{acknowledgements}
This work was written in part during stays of TJLC at ESO in Santiago and Garching. The hospitality of ESO is acknowledged with pleasure. We benefited from many discussions with  M. Chernyakova, J. Kirk and L. Woltjer. The electron Compton heating was specifically discussed with A. Benz.
\end{acknowledgements}

\end{document}